\documentstyle[prl,aps]{revtex}
\twocolumn

\begin{document}
\title{SUPERFLUID - BOSE-GLASS TRANSITION IN WEAKLY DISORDERED\\
COMMENSURATE ONE-DIMENSIONAL SYSTEM}

\author {
Boris V. Svistunov}
\address{
Russian Research Center "Kurchatov Institute", 123182 Moscow, Russia}

\maketitle
\begin{abstract}
We study the effect of commensurability (integer filling factor) on
the superfluid (SF) - Bose-glass (BG) transition in a one-dimensional
disordered system in the limit of weak disorder, when the effect is
most pronounced and, on the other hand, may be traced via the 
renormalization-group analysis. The equation for the SF-BG
phase boundary demonstrating the effect of disorder-stimulated 
superfluidity implies that the strength of disorder sufficient to 
restore superfluidity from Mott insulator (MI)
is much larger than that enough to turn MI into BG. Thus we provide an
explicit proof of the fact that at arbitrarily small disorder the SF
and MI phases are always separated by BG.
\vspace{0.2cm} \\
PACS numbers: 05.30.Jp, 64.60Ak, 67.40.-w
\vspace{0.5cm}
\end{abstract}

The problem of superfluid - insulator transitions in commensurate and 
disordered bosonic systems has attracted a lot of interest in
the past decade (for an introduction see Refs.\ \cite{MHL,GS,FWGF}).
A special part in the analysis of the problem is played by one-dimensional 
(1D) systems, where along with the exact finite-size numerical treatment
(quantum Monte Carlo \cite{BSZ,MTU} and exact diagonalization \cite{R,KPS}) 
there exists macroscopic renormalization-group (RG) description 
\cite{GS}. [Recently it was demonstrated that the 
two are in an excellent agreement with each other at least in the
case of SF-MI transition \cite{KS}.] Note also a remarkable success
of density-matrix RG study of 1D bosonic Hubbard model \cite{PPKR}.

Despite the fact that now the understanding of superfluid - insulator
transitions is close to being complete, especially in 1D, there is
one point, however, which remains somewhat unclear up to now. That is
the question of superfluid - insulator transition in a weakly disordered
commensurate system with integer filling factor. From general 
considerations one would expect here that SF and MI phases are always
separated by BG phase \cite{FWGF,FM}, while a number of approximate 
and exact (but finite-size!) methods suggest that small disorder
is not relevant and there occurs a direct 
transition from SF to MI \cite{KTC,SR,PPKR}. In this paper we consider 
this problem in 1D taking advantage of the asymptotically exact RG 
treatment.

Our main result is the demonstration of the existence of the BG phase 
at arbitrarily weak disorder. We provide the proof of this fact in a
'direct' and in an 'enhanced' forms. By 'direct' form we mean tracing
the renormalization picture (in the superfluid phase, where the RG 
equations are asymptotically exact) with the observation that the
commensurate scenario, taking place at smaller scales of distance,
is destroyed at larger distances by sufficiently strong disorder,
in favor of the universal disordered scenario. [Incidentally, this
accounts for the effect of disorder-stimulated superfluidity.] So
that commensurability turns out to be irrelevant to any 
second-order-type transition arising with decreasing of the strength
of disorder. The 'enhanced' proof does not rely on any assumption
concerning the type of the transition leading to the destruction
of superfluidity. We simply take the phase boundary for
superfluid - insulator transition following from the 'direct' 
treatment as a lower boundary for the strength of disorder 
sufficient to support superfluidity, without specifying the phase
adjacent to SF. Then we compare the result with that for the MI 
phase boundary, following from the exact relation between the minimal 
amount of disorder destroying MI and the insulating gap of the pure 
system \cite{FM}. The comparison shows that, in relative units, the 
separation between the boundaries is ever increasing with vanishing 
disorder, thus revealing the presence of BG between MI and SF.

The RG descriptions of the superfluid-to-insulator transitions in 1D
were initially constructed in terms of the effective action for the
long-wave density deviations \cite{GS,FWGF}, employing Haldane's 
representation of 1D bosons \cite{Hald}. For our purposes, however, 
it is more convenient to use a dual treatment in terms of the 
effective action for the phase field \cite{KPPS}. This language 
directly takes into account effects of the collective backscattering 
off either commensurate or disordered potential, the corresponding 
instantons being just vortices in the phase
field in (1+1) dimensions. As a result, the descriptions for the
commensurate and disordered cases turn out to be very close
to each other. The only difference is that in the disordered case
one deals with 'vertical' vortex-antivortex pairs (the spatial 
coordinates of vortex, $x_1$, and antivortex, $x_2$, coincide up to
some microscopic length), because contributions of non-vertical pairs 
are averaged out due to the random phase associated with the pair:
\begin{equation}
\gamma (x_1,x_2) = -2 \pi \int_{x_1}^{x_2} n_0(x) \: dx \; ,
\label{gam1}
\end{equation}
where $n_0(x)$ is the expectation of the density at the point $x$
\cite{KPPS}. In the weak-disorder limit $n_0(x)$ is related to the
disordered potential, $\tilde{W}(x)$, perturbatively:
\begin{equation}
n_0(x) =n^{(0)}(x) + 
\int dx^{\prime} \: S(x^{\prime},x) \tilde{W}(x^{\prime})\; ,
\label{n_0}
\end{equation}
where $n^{(0)}(x)$ and $S(x^{\prime},x)$ are the density 
expectation and static susceptibility for the pure system.
\begin{equation}
S(x^{\prime},x) = i \int_0^{\infty} dt \;
\langle \: \left[ n(x^{\prime},t), n(x,0) \right] \: \rangle \; ,
\label{S}
\end{equation}
$n$ is the density operator. Substituting (\ref{n_0}) into 
(\ref{gam1}) and taking into account that in the integer-filling 
case the contribution from $n^{(0)}(x)$ leads only to some 
irrelevant renormalizations \cite{KPPS}, we get
\begin{equation}
\gamma (x_1,x_2) = -2\pi \int_{x_1}^{x_2} dx \: Q(x) \tilde{W}(x) \; ,
\label{gam2}
\end{equation}
where
\begin{equation}
Q(x) = \int dx^{\prime} \: S(x,x^{\prime}) 
\label{Q}
\end{equation}
is a periodic (commensurate) function. [In Eq.(\ref{gam2})  
$\mid x_2 - x_1 \mid$ is meant to be much greater than the 
microscopic static correlation radius characterizing the decay of 
$S(x^{\prime},x)$ as a function of $\mid x - x^{\prime} \mid$.]

For definiteness, we take $\tilde{W}(x)$ to be bounded,
\begin{equation}
-W/2 \: \leq \: \tilde{W}(x) \: \leq \: W/2 \; ,
\label{W}
\end{equation}
and symmetric with respect to the change of the sign.

Introducing the radius of a vortex pair
\begin{equation}
R = \sqrt{(x_2-x_1)^2 + (y_2-y_1)^2} \; ,
\label{R}
\end{equation}
$y_1$ ($y_2$) being the coordinate of the vortex (antivortex) along
the imaginary-time axis ($y=c \tau$, $c$ is the sound velocity, 
$\tau$ is the imaginary time),
we see from Eq.(\ref{gam2}) that the mean-root-square value of 
the phase associated with a pair of the radius $R$ is
\begin{equation}
\gamma_R \, \sim \, W \sqrt{R} \; .
\label{gam_R}
\end{equation}
[From now on we treat $W$ and $R$ as dimensionless quantities,
assuming them to be measured in units of characteristic 
interparticle energy and distance, respectively.] 
While $\gamma_R \ll 1$, the
disorder is irrelevant and the renormalization picture follows
the scenario of the pure commensurate case. At the scale
\begin{equation}
R_* \, \sim \, 1/W^2 \; ,
\label{R_*}
\end{equation}
corresponding to $\gamma_{R_*} \sim 1$, the renormalization picture
crosses over to the disordered scenario which becomes well-developed
at $R \gg R_*$, when only the vertical pairs 
($\, \mid y_2-y_1 \mid \, \gg \, \mid x_2-x_1 \mid \, \sim R_*$) 
contribute.
The scale $R_*$ itself plays negligible role in the renormalization,
since near the critical point in 1D the total contribution comes
from exponentially large range of distances. We thus can simply
sew together the known RG descriptions \cite{GS} 
(see also \cite{FWGF,KPPS}) for the purely commensurate and the purely
disordered cases:
\begin{equation}
\frac{dK}{d\lambda} = w^2 \; , \; \; 
\frac{dw}{d\lambda} = \left( 2 - 1/K \right) w  \; \; \; \; \;
(\lambda \leq \lambda_*) \; ; 
\label{RG1}
\end{equation}
\begin{equation}
\frac{dK}{d\lambda} = K w^2 \; , \; \;
\frac{dw}{d\lambda} = \left(3/2 - 1/K \right) w  \; \; \;
(\lambda \geq \lambda_*) \; .
\label{RG2}
\end{equation}
Here $\lambda = \ln R$, $\lambda_* = \ln R_*$; 
$K = \left[ \pi \sqrt{\Lambda_s \kappa} \, \right]^{-1}$ 
($\Lambda_s$ is the superfluid stiffness, $\kappa$ is 
the compressibility). The quantity $[w(\lambda)]^2$ is proportional 
to the number of vortex pairs of the size $\, \sim R$ in the area 
$\, \sim R^2$ of the $xy$-plane \cite{KPPS}, and is much less than 
unity in the region of applicability of Eqs.(\ref{RG1})-(\ref{RG2}).
Strictly speaking, $w(\lambda)$ experiences a jump at 
$\lambda = \lambda_*$, but this is not important for our 
consideration.

The critical value of $K(\lambda = \infty)$ for the destruction of
superfluidity (divergency of $K(\lambda)$ as 
$\lambda \to \infty$) is $1/2$ in the commensurate case 
\cite{Col,Hald,GS} and $2/3$ in the disordered one \cite{GS}.
Hence, if $K(\lambda_*) < 2/3$, then it may turn out that
$K(\infty) < 2/3$ also, and the system is in the superfluid phase
though in the pure case it would be Mott insulator. In the 
weak-disorder limit one may neglect the difference between 
$K(\infty)$ and
$K(\lambda_*)$, provided the system is in SF phase [that is 
neglect the renormalization of $K$ due to disorder]. The critical
condition for the SF-BG transition then reads
\begin{equation}
K(\lambda_*) = 2/3 \; .
\label{crit}
\end{equation}
To reveal the behavior of $K(\lambda_*)$ as a function of the
Hamiltonian parameters consider the solution of the system 
(\ref{RG1})-(\ref{RG2}) near the critical value $K=1/2$ from
the Mott-insulator side:
\begin{equation}
K = \frac{1}{2} + a \tan \left[ 4a \lambda - \pi /2 \right] \; \; \; \;
(\mid K - \frac{1}{2} \mid \: \ll 1) \; .
\label{K}
\end{equation}
[A free constant in the argument of the tangent is fixed to be
$- \pi /2$ (up to a small uncertainty on the order $a \ll 1$)
by the condition that at the microscopic scale $\lambda \sim 1$
the parameter $K(\lambda)$ should be noticeably smaller than its
mesoscopic value $1/2$.] 
Near the critical point the quantity $a$ depends on the parameters 
of the Hamiltonian of the pure system in the square-root fashion: 
\begin{equation}
a \propto \sqrt{p/p_0 -1} \; .
\label{a}
\end{equation}
Here $p$ is any parameter of the Hamiltonian, the value $p=p_0$ 
corresponding to the point of SF-MI transition in the pure system 
($p > p_0$ in the Mott phase). For example, one may choose
$p$ to be the strength of interparticle interaction.

From Eq.(\ref{K}) it is seen that $K(\lambda)$ does not differ 
essentially from $1/2$ until $\lambda$ comes closely to the point
$\pi /4a$. At this point $K$ diverges and we thus conclude that to 
a good accuracy 
\begin{equation}
\lambda_* = \frac{\pi}{4a} \; .
\label{lambda_*}
\end{equation}
A comment is in order here concerning the usage of Eq.(\ref{K}) 
at $K=2/3$, i.e. out of the region of its formal applicability.
Actually, at this value of $K$ Eq.(\ref{K}) yields only a rough
order-of-magnitude estimate. This roughness, however, is
essentially compensated by the sharp behavior of $K(\lambda)$
in the vicinity of the point $\lambda = \pi /4a$. So that the 
relative uncertainty of the value of $\lambda_*$ is easily
estimated to be on the order $a \ll 1$.

Combining Eqs.\ (\ref{R_*}), (\ref{a}), and (\ref{lambda_*}) we 
obtain the equation for the phase boundary between superfluid and
Bose glass [$W=W_{\mbox{\scriptsize SF-BG}}(p)$]:
\begin{equation}
W_{\mbox{\scriptsize SF-BG}}(p) \: \sim \:
\exp \left( - \frac{\pi}{8b \sqrt{p/p_0-1}} \right) \; ,
\label{SF_BG}
\end{equation}
where $b$ is some constant. Eq.(\ref{SF_BG}) demonstrates the effect
of disorder-stimulated superfluidity in the vicinity of the SF-MI
transition of the pure commensurate system, which was observed
previously in numerical studies \cite{KTC,PPKR}. Note a very sharp 
behavior of the critical value of $p$ for the SF-BG transition as a 
function of the strength of disorder.

To provide the 'enhanced' proof of the existence of the BG in the
weak disorder limit we compare Eq.(\ref{SF_BG}) with the
equation for the phase boundary between MI and a gapless
phase (which subsequently will be identified with BG; until this
is done the notations MI-BG and SF-BG should be understood as a
convention). To obtain the latter equation one can take advantage
of the exact relation taking place at the MI phase boundary
\cite{FM}:
\begin{equation}
\Delta = W \; ,
\label{Delta}
\end{equation}
where $\Delta$ is the insulating gap of the pure system. This
relation follows from the fact that the infinitesimal destruction
of the MI is due to the most favorable exchange of particles
between exponentially rare Lifshitz's regions, where local
configurations of disorder mimic uniform potential of the
value $W/2$ or $-W/2$ \cite{FWGF,FM}. An estimate for $\Delta$ 
one can obtain from the relation $\Delta \sim 1/R_*$ \cite{FWGF}. 
So we have
\begin{equation}
W_{\mbox{\scriptsize MI-BG}}(p) \: \sim \:
\exp \left( - \frac{\pi}{4b \sqrt{p/p_0-1}} \right) 
\label{MI_BG}
\end{equation}
and see that
\begin{equation}
\frac{W_{\mbox{\scriptsize SF-BG}}(p)}{W_{\mbox{\scriptsize MI-BG}}(p)} 
\: \sim \: \exp \left( \frac{\pi}{8b \sqrt{p/p_0-1}} \right) 
\; \gg 1 \; .
\label{ratio}
\end{equation}
Hence, in a weakly disordered commensurate 1D system the strength 
of disorder sufficient to restore superfluidity is exponentially 
larger than that enough to turn Mott insulator into a gapless phase.
This immediately proves the existence of BG between SF and MI,
and identifies (\ref{SF_BG}) and (\ref{MI_BG}) with corresponding
phase boundaries.
Ultimately, this effect is due to the fact that $\Delta$ scales
like $1/R_*$, while the critical amplitude of disorder supporting
superfluidity is $\sim 1/ \sqrt{R_*}$.

Let us now discuss recent results of numerical density-matrix 
RG study of 1D commensurate disordered bosonic Hubbard
model \cite{PPKR}. In the weakly disordered case the authors observe
a direct transition from SF to MI, in contrast to the case of a 
substantial disorder where BG intervenes between SF and MI, and to
our asymptotically exact (at $W \to 0$) analytical results.

It is known, however, that in the limit of weak disorder it is
difficult to distinguish BG from MI by a finite-size scaling,
because the difference is associated with exponentially rare
Lifshitz's regions \cite{FWGF,FM}. If the size of the system is 
not large enough,
these regions are simply absent with probability close to unity,
and the scaling analysis reveals nothing. [Maximal size of the 
system in Ref.\cite{PPKR} is of the order of $100$ sites.]

Therefore, one may suspect that the transition observed in 
Ref.\cite{PPKR} at a small disorder is actually the SF-to-BG one. 
In this connection it is interesting to compare the numerical data
with Eq.(\ref{SF_BG}). Unfortunately, the phase diagram in 
Ref.\cite{PPKR} contains only few points in the weak-disorder
region, too little to make possible an unambiguous fitting. 
However, one may check a non-trivial qualitative prediction of 
Eq.(\ref{SF_BG}), the requirement that the strength of disorder
on the SF-BG phase boundary be much larger than the Mott gap in
the absence of disorder. The data clearly demonstrate this feature.
At any of the three presented points on the superfluid - insulator 
phase boundary in the weak-disorder region the strength of disorder 
is at least an order of magnitude larger than the insulating gap.

We can allow also for the effect of a small deviation from 
commensurability. To this end we should add to the right-hand side of 
Eq.(\ref{gam2}) the term $2 \pi \delta n^{(0)} (x_1-x_2)$, where 
$\delta n^{(0)}$ is the deviation of the density from the commensurate 
value. This term is negligible for a vortex pair of a radius
much less than
\begin{equation}
R_c \sim \: \mid \delta \nu \mid^{-1}
\label{R_c}
\end{equation}
($\delta \nu$ is the deviation of the filling factor), and leads to
the cancellation of contributions of pairs with radii
$R_c < R < R_*$, provided $R_c < R_*$. At $R_c > R_*$
this term plays no role at all. Hence, at $R_c < R_*$, we just
need to replace $\lambda_*$ by $\ln R_c$ in Eq.(\ref{RG1}) and
set $K(\lambda_*) = K(\ln R_c)$ as the initial condition for
Eq.(\ref{RG2}). Consequently, for the SF-BG phase boundary
we obtain the equation
\begin{equation}
\exp \left( - \frac{\pi}{4b \sqrt{p/p_0-1}} \right)
\: \sim \: \max \left\{ W^2, \mid \delta \nu \mid \right\} \; .
\label{cusp}
\end{equation}
Note that in the variables $p$ and $\delta \nu$ (at fixed $W$)
the phase boundary has an independent of $W$ cusp-like shape
except for the tip of the cusp cut off by disorder.

A remark is in order here concerning fractional filling factors.
The fact that the point of SF-MI transition of a pure commensurate
system turns out to be the limiting point for both SF-BG and BG-MI
phase boundaries of the disordered system is characteristic of only
integer filling factors, because only in this case the critical
value of the parameter $K$ for the pure system, $K=1/2$, is less
than that of disordered system, $K=2/3$. In the vicinity of
SF-MI transition at a fractional filling, where the transition 
occurs at $K=q^2/2$ ($q$ is the denominator of the filling factor) 
\cite{Col,Hald}, an infinitesimal disorder takes the system
away from the superfluid region.

Summarizing, we have presented a description of the superfluid - 
Bose-glass phase boundary in 1D weakly disordered system
near the point of superfluid - Mott-insulator transition of
the pure system with integer filling factor. The results
demonstrate the effect of stimulation of superfluidity by
disorder, previously observed numerically. In the limit of weak 
disorder the strength of disorder sufficient to restore superfluidity 
from Mott insulator turns out to be much larger than that enough 
to turn Mott insulator into Bose glass. This provides an explicit 
proof of the fact that at arbitrarily small disorder the Bose glass 
always intervene between superfluid and Mott phases.

The author is grateful to V.Kashurnikov, N.Prokof'ev, and
J.Freericks for valuable discussions.

This work was supported by the Russian Foundation for Basic
Research (under Grant No. 95-02-06191a) and by the Grant 
INTAS-93-2834-ext [of the European Community].

\end{document}